\documentstyle[12pt,fleqn]{article}
\begin{document}
\setlength{\unitlength}{1mm}
\title{\hfill {\small  Alberta Thy 20-96} \\
\hfill {\small  July 1996} \\
\vskip 2cm
Statistical Origin of Black Hole Entropy in Induced Gravity}
\author{\\
V.P. Frolov$^{1,2,3}$,
D.V. Fursaev$^{1,4}$, and
A.I. Zelnikov$^{1,3}$ \date{}}
\maketitle
\noindent  {
$^{1}${ \em
Theoretical Physics Institute, Department of Physics, \ University of
Alberta, \\ Edmonton, Canada T6G 2J1}
\\ $^{2}${\em CIAR Cosmology Program}
\\ $^{3}${\em P.N.Lebedev Physics Institute,  Leninskii Prospect 53,
Moscow
117924, Russia}
\\ $^{4}${\em Joint Institute for
Nuclear Research, Laboratory of Theoretical Physics, \\
141 980 Dubna, Russia}
\\
\\
e-mails: frolov, dfursaev, zelnikov@phys.ualberta.ca
}
\bigskip

\begin{abstract}
The statistical-mechanical origin of the Bekenstein-Hawking entropy $S^{BH}$ in
the induced gravity is discussed. In the framework of the induced gravity
models the Einstein action arises as the low energy limit of the effective
action of quantum fields. The induced gravitational constant is determined by
the masses of the heavy constituents.  We established the explicit relation
between statistical entropy of constituent fields  and black hole entropy
$S^{BH}$.
\end{abstract}

\bigskip

{\it PACS number(s): 04.60.+n, 12.25.+e, 97.60.Lf, 11.10.Gh}

\baselineskip=.6cm

\newpage

\noindent
\section{Introduction}

It is well known \cite{Beke:72},\cite{Hawk:75} that a black hole in the
Einstein gravity  behaves as a thermodynamical system with
the entropy
	\begin{equation}\label{1.1}
	{S}^{BH} ={ 1 \over 4 } {~{\cal A}^H \over  G}~~~,
	\end{equation}
where ${\cal A}^H$   is   the   area   of the event  horizon
and $G$ is the Newton constant. The Bekenstein-Hawking entropy
${S}^{BH}$ is the physical quantity which can be measured  in
(gedanken) experiments by making use of
the first law of thermodynamics which can be represented in the form
\cite{GiHa:76}:
	\begin{equation}\label{1.2}
	dF^{H}=-{S}^{BH}dT^{H}~~~.
	\end{equation}
Here $M^{H}$ is mass of a black hole, $T^{H}=(8\pi M^{H})^{-1}$ is
its Hawking temperature and $F^{H}=M^{H}-T^{H}S^{BH}$ is the free
energy.

Recently a lot of attempts has been made to provide the statistical-mechanical
foundation of the black hole thermodynamics and in particular to relate
$S^{BH}$ with counting the internal degrees of freedom of a black hole.
One of the proposed ideas is to relate the dynamical degrees of freedom of a
black hole with its quantum excitations
\cite{Hoof:85}-\cite{FroNo}. However, this approach meets the evident
difficulty because
the Bekenstein-Hawking entropy arises at the tree-level while the entropy of
quantum excitations is a one-loop quantity
\cite{Frol:95}-\cite{FFZ2}. This difficulty might be overcome if the gravity
itself arises as the result of quantum effects, so that the metric $g_{\mu\nu}$
is a collective variable and the general relativity is a low-energy effective
theory \cite{Jacobson}.

One of the possible realizations of this idea is the  theory of the induced
gravity proposed by Sakharov \cite{Sakh}. According to this idea the background
 fundamental theory is described by the action $I[\varphi_i,g_{\mu\nu}]$ of
fields $\varphi_i$ propagating in the external geometry with the metric
$g_{\mu\nu}$. By averaging over the constituent fields $\varphi_i$ one gets the
dynamical effective action for the metric $g_{\mu\nu}$
\begin{equation}\label{1.3}
\exp(-W[g_{\mu\nu}])=\int {\cal D}\varphi_i \exp(-I[\varphi_i,g_{\mu\nu}])~~~.
\end{equation}
This approach resembles the well-known approach in the solid state physics when
instead of
variables describing the oscillations of the atoms of the lattice one uses new
collective variables describing the phonon field. It is important that the
thermodynamical characteristics of a solid state in the low energy regime, say
at the low temperatures, can be expressed by using only the spectrum of the
phonon excitations.

The idea of this paper is to relate the Bekenstein-Hawking entropy with the
statistical mechanics of the ultraheavy
constituents that induce the gravity in the low energy limit of the theory.  We
shall illustrate it by direct calculations in a simple class of induced gravity
models. Our result is an explicit representation for the
Bekenstein-Hawking entropy
\begin{equation}\label{1.4}
S^{BH}=-\sum_{i}\mbox{Tr}~\rho_i\ln\rho_i-\sum_{s}
2\pi\xi_s\int_{\Sigma}<\hat{\phi}^2_s>d\sigma
\end{equation}
in terms of the statistical-mechanical  entropies $-\mbox{Tr}\rho_i\ln\rho_i$
of all heavy
constituents
and the average of the square of the field operators on the horizon $\Sigma$.
The latter quantity appears for the fields which have  coupling terms with the
Riemann tensor
in the Lagrangian. Such interactions  are specific, for instance, for gauge
fields. In our model we demonstrate an appearance of these terms for the scalar
fields $\phi_s$ with nonminimal couplings $\xi_s R\phi_s^2$ with the scalar
curvature.

The arguments in favour of the idea that
the black hole entropy can be explained in the framework of the induced gravity
were first given by T. Jacobson \cite{Jacobson}. In our paper we analyse the
statistical-mechanical origin between $S^{BH}$ in such theories
and establish explicit relation (\ref{1.4}) of $S^{BH}$ and the entropies of
the constituent fields.

\section{Induced gravity: a simple model}
\setcounter{equation}0

One of the simplest models that can be used to illustrate the idea of the
induced gravity consists of $N_s$ scalar fields $\phi_s$ with the classical
actions
\begin{equation}\label{i1}
I[\phi_s,g_{\mu\nu}]=\frac 12\int_{\cal M}dV\left[\left(\nabla \phi_s\right)^2
+\left(m_s^2+\xi_s R\right)\phi_s^2\right]+
\xi_s\int_{\partial {\cal M}}dv~K\phi^2_s~~~,
\end{equation}
and $N_d$ fields described by the Dirac fermions with the actions
\begin{equation}\label{i2}
I[\psi_d,g_{\mu\nu}]=\int_{\cal M}dV~\bar{\psi}_d\left(i\gamma^\mu\nabla_\mu+
m_d\right)\psi_d~~~.
\end{equation}
Here, $dV$ and $dv$ are the volume elements of the background space ${\cal M}$
and its boundary $\partial {\cal M}$. Indices $s=1,..,N_s$ and $d=1,..,N_d$
enumerate the scalar and Dirac fields, correspondingly.
It is important for our purpose to introduce into  the classical actions
(\ref{i1}) the non-minimal couplings with the curvature $R$. In order
to have a consistent variational principle  for the metric $g_{\mu\nu}$ one
must also add  the boundary term to $I[\phi_s,g_{\mu\nu}]$. The variational
procedure, which corresponds to this  action, fixes only the metric on
$\partial {\cal M}$ and keeps
free the extrinsic curvature $K$ of the boundary.

The effective gravitational action $W[g_{\mu\nu}]$ is defined by Eq.
(\ref{1.3}) where the classical action for the constituent fields is the sum
\begin{equation}\label{i8}
I[\varphi_i,g_{\mu\nu}]=\sum_s I[\phi_s,g_{\mu\nu}]+
\sum_d I[\psi_d,g_{\mu\nu}]~~~.
\end{equation}
We assume that the fields (that can have different masses and the coupling
constants $\xi_s$) obey the following constraints:
\begin{equation}\label{i3}
c_1=N_s-4 N_d=0~~~,
\end{equation}
\begin{equation}\label{i4}
c_2=N_s+2N_d-6\sum_{s}\xi_s=0~~~,
\end{equation}
\begin{equation}\label{i5}
c_3=\sum_{s} m_s^2-4 \sum_{d} m_d^2=0~~~,
\end{equation}
\begin{equation}\label{i6}
c_4=\sum_{s} m_s^2 (1-6\xi_s)+2\sum m^2_d=0~~~,
\end{equation}
\begin{equation}\label{i6'}
c_5=\sum_{s} m_s^4-4 \sum_{d} m_d^4=0~~~,
\end{equation}
\begin{equation}\label{i7}
c_6=\sum_{s} m_s^4\ln m_s^2 -4 \sum_{d} m_d^4\ln m_d^2=0~~~.
\end{equation}
In 4-dimensional space-time the first five constraints ensure the ultraviolet
finiteness of the induced Newton and cosmological constants ($G$ and $\Lambda$,
respectively).
The last constraint (\ref{i7}) is imposed for the $\Lambda$-constant to vanish
so that the induced gravity theory
possesses the vacuum black hole solutions. As it is seen from Eq.(\ref{i4}) the
finite gravitational constant is impossible in the models like (\ref{i8})
without curvature couplings, i.e.
when all $\xi_s=0$.

The main preposition of the induced gravity is that at least some of the
constituent fields possess masses comparable with the Planckian mass $m_{Pl}$.
The low energy limit of the theory corresponds to the regime when the curvature
radius $L$ of the space-time ${\cal M}$ is much greater than the Planck length
$m_{Pl}^{-1}$.  In this limit the functional $W[g_{\mu\nu}]$
can be approximated by the Euclidean Einstein action
\begin{equation}\label{i9}
W[g_{\mu\nu}]\simeq -{1 \over 16\pi G}\left(\int_{\cal M}dVR+2\int_{\partial
{\cal M}}dv K\right)
\end{equation}
where the Newton constant $G$ is the function of the parameters of the
constituents
\begin{equation}\label{i10}
{1 \over G}={1 \over 12\pi} \left(\sum_{s}(1-6\xi_s)~ m_s^2 \ln m_s^2+2\sum_{d}
m^2_d\ln m_d^2\right)~~~.
\end{equation}
The value of this  constant is dominated by the masses of the heavy
constituents $G \sim m_{Pl}^{-2}$.

Now several remarks are in order how to obtain Eq.(\ref{i9}). The one-loop
expression for the action $W[g_{\mu\nu}]$ in the Schwinger-DeWitt
representation
is
\begin{equation}\label{i11}
W[g_{\mu\nu}]=\sum_{s}\left( W_s+\xi_s\int_{\partial {\cal M}}dvK
<\hat{\phi}_s^2>\right)+\sum_{d} W_d~~~,
\end{equation}
where $W_s$ and $W_d$ are the effective actions of a
single field $\varphi_i$
\begin{equation}\label{i12}
W_i=-{\eta_i \over 2} \int_{\delta}^{\infty}{ds \over s} e^{-m_i^2s}~
\mbox{Tr}~e^{-s\triangle_i }~~~,~~~i=s,d~~~,
\end{equation}
and the factors $\eta_s=1$, $\eta_d=-1$ correspond to the different statistics
of the fields $\phi_s$ and $\psi_d$. The wave operators read
\begin{equation}\label{i13}
\triangle_s=-\nabla^2+\xi_s R~~~,
{}~~~\triangle_d=-(\gamma^\mu\nabla_\mu)^2=-\nabla^2+
\frac 14 R~~~.
\end{equation}
The boundary term in (\ref{i11}) appears in the one-loop approximation in the
perturbation theory (see for the details \cite{BaSol}) and the field average is
defined as
\begin{equation}\label{i14}
<\hat{\phi}_s^2(x)>=\int_{\delta}^{\infty}ds~ e^{-m_i^2s}~ <x|e^{-\triangle_i
s}|x>~~~.
\end{equation}
In Eqs.(\ref{i12}) and  (\ref{i14}) the parameter $\delta$ is the ultraviolet
regulator that must be introduced in order to
make finite the integral for each separate field in the limit $s\rightarrow
0$.
In intermediate computations it is possible to use other type of the
ultraviolet regularizations, but  in the induced gravity models after the
regularization is removed the physical quantities are well defined and do not
depend on the chosen regularization scheme.

In the low energy limit the main contribution in the integrals
(\ref{i12}) and (\ref{i14}) comes from  the small values of the proper-time
parameter $s\sim m_{Pl}^{-2}$, so that one can use the asymptotic expansion of
the heat kernel
\begin{equation}\label{i15}
\mbox{Tr}~e^{-\triangle_i s}\simeq {1 \over (4\pi s)^{D/2}}\left((A_0)_i+
(A_{1/2})_i~s^{1/2}+
(A_1)_i ~s+...\right)~~~,
\end{equation}
with $D$ standing for the dimension of the background geometry.
The first heat coefficients are well-known:
\begin{equation}\label{i16}
(A_0)_s=\int_{\cal M}dV~~~,~~~(A_0)_d=4\int_{\cal M}dV~~~,
\end{equation}
\begin{equation}\label{i17}
(A_1)_s=\left(\frac 16-\xi_s\right)\int_{\cal M}dV R+\frac 13 \int_{\partial
{\cal M}}dv K~~~,~~~
(A_1)_d=-2\left.(A_1)_s\right|_{\xi_s=0}~~~.
\end{equation}
The coefficients $(A_{1/2})_i$ depend on the imposed boundary conditions and
they are proportional to the volume of
$\partial {\cal M}$. For this reason in the variational procedure, where the
metric of $\partial {\cal M}$ is fixed, the coefficients
$(A_{1/2})_i$ give the constant contributions to the effective action, and
therefore they can be neglected.

If we approximate now the heat kernels in (\ref{i11}), (\ref{i12}) and
(\ref{i14}) by the asymptotic expansions (\ref{i15}), then we get the
decomposition of $W[g_{\mu\nu}]$ in the curvature  with the small expansion
parameter $(m_{Pl}L)^{-1}$. Eq. (\ref{i9}) is obtained  when all terms higher
then the first order in curvature $R$ are neglected. The gravitational constant
reads:
\begin{equation}\label{i18}
\frac 1G={1 \over 12\pi}\left[\sum_s m_s^{D-2}(1-6\xi_s)
{}~\Gamma\left(1-D/2,m_s^2\delta\right)+2
\sum_d m_d^{D-2}
\Gamma\left(1-D/2,m_d^2\delta\right)\right]
\end{equation}
where
\begin{equation}\label{i19}
\Gamma(z,\sigma)=\int_{\sigma}^{\infty}x^{z-1}e^{-x}dx
\end{equation}
is incomplete gamma function. In four dimensions if the constraints
(\ref{i4}), (\ref{i6}) hold the $G$ remains finite in $D=4$ in the limit
$\delta=0$ and  is given by Eq.(\ref{i10}).

Note that the $R^2$-terms that are neglected here are ultraviolet divergent and
so they must be either renormalized or made finite by adding further
restrictions and complicating the model. However we will not dwell on this
issue because it is not related with the Bekenstein-Hawking entropy (\ref{1.1})
which appears in the Einstein theory of gravity.

\section{Induced entropy: on and off shell}
\setcounter{equation}0

The black hole entropy can be derived for the induced action by  standard
methods. One may use the York \cite{York:86},\cite{BBWJ} formulation of the
canonical ensemble for black holes
inside a spherical cavity of a radius $r_B$ with the fixed temperature $T$ on
its surface. The free energy $F$ of the system is defined in terms of the
effective action
$W[g_{\mu\nu}]$, taken on the Euclidean black hole instanton ${\cal M}$,  as
$F=T~W[g_{\mu\nu}]$. Then
the second law of thermodynamics (\ref{1.2}) gives for the entropy $S^{BH}$ the
Bekenstein-Hawking expression (\ref{1.1}). Such a definition of the entropy is
called thermodynamical.
It considers only equilibrium changes of the system. This is equivalent to the
requirement that
the effective action is always taken on-shell, i.e for the regular
black hole instanton that is a solution of the vacuum Euclidean Einstein
equations.

Contrary to the thermodynamical approach the statistical-mechanical one, as it
was pointed out by many authors
\cite{Frol:95}-\cite{FFZ2}, requires the off-shell computations. Let $\beta$ be
the period of the Euclidean time-coordinate $\tau$ of ${\cal M}$, connected
with the temperature $T$ on the boundary $\partial {\cal M}$ as
$\beta=g_{\tau\tau}^{-1/2}(r_B)T^{-1}$ where $g_{\tau\tau}$ is the time
component of
the metric. For arbitrary $\beta$ and  fixed mass $M^H$  the
black hole instanton has the conical singularity at the horizon $\Sigma$ with
the conical deficit angle $2\pi(1-\beta/\beta_H)$. Here, $\beta_H^{-1}$ is the
Hawking temperature ($\beta_H=8\pi M^H$ for the Schwarzschild solution).
The off-shell entropy $S^{CS}$ in the conical-singularity method is defined as
\begin{equation}\label{e1}
S^{CS}(\beta)=\left(\beta {\partial \over \partial
\beta}-1\right)W^{CS}[g_{\mu\nu},\beta]~~~,
\end{equation}
where $W^{CS}[g_{\mu\nu},\beta]$ is the action on the singular instanton
($W^{CS}[g_{\mu\nu},\beta_H]=W[g_{\mu\nu}]$) and the derivative with respect to
 $\beta$ is taken under fixed $g_{\mu\nu}$.
Let us compute the value of $S^{CS}(\beta)$ in the induced gravity and show
that at the Hawking temperature this quantity coincides with the
Bekenstein-Hawking entropy $S^{BH}$.

First, it is important to note that in a space with the conical singularities
the total scalar curvature $\bar{R}$ in addition to its standard regular part
$R$  has a delta-function contribution concentrated on $\Sigma$
\begin{equation}\label{e2}
\bar{R}=R+4\pi(1-\beta/\beta_H)\delta_{\Sigma}~~~.
\end{equation}
Therefore the classical action $I^{CS}[\phi_s,g_{\mu\nu}]$ for a scalar field
differs from the action (\ref{i1}) on a regular manifold by a term on the
horizon
\begin{equation}\label{e3}
I^{CS}[\phi_s,g_{\mu\nu}]=I[\phi_s,g_{\mu\nu}]+2\pi\xi_s
(1-\beta/\beta_H)\int_{\Sigma}d\sigma \phi^2_s~~~,
\end{equation}
($d\sigma$ is the area element of $\Sigma$). The spinor action (\ref{i2}) does
not change.
According to this observation the one-loop effective action in the conical
singularity method $W^{CS}$ can be written as \cite{Solod}
\begin{equation}\label{e4}
W^{CS}[g_{\mu\nu},\beta]=W[g_{\mu\nu},\beta]+
\sum_{s}2\pi\xi_s(1-\beta/\beta_H)\int_{\Sigma}d\sigma
<\hat{\phi}^2_s>_{\beta}
\end{equation}
where $W[g_{\mu\nu},\beta]$ and $<\hat{\phi}^2_s>_{\beta}$ are given by  Eqs.
(\ref{i12}) and (\ref{i14}) in terms of the heat kernel operator on the
singular manifold. To calculate these quantities one can use the asymptotic
heat kernel expansion (\ref{i15}) with the modified
integer index coefficients. In particular
the first scalar \cite{DF},\cite{CKV} and spinor \cite{Kabat},\cite{FM96}
coefficients
$(A_1^{CS})_i$ read
\begin{equation}\label{e5}
(A_1^{CS})_i=(A_1)_i+(A_{1,\beta})_i~~~,~~~i=s,d~~~,
\end{equation}
\begin{equation}\label{e6}
(A_{1,\beta})_s={\pi \over 3}\left({\beta_H \over \beta}-
{\beta \over \beta_H}\right){\cal A}^H~~~,
{}~~~(A_{1,\beta})_d=-2(A_{1,\beta})_s~~~,
\end{equation}
where ${\cal A}^H=\int_\Sigma d\sigma$ is the area of the black hole horizon.
Note that $(A_1)_i$ are given by  integrals (\ref{i17}) and they are
proportional to $\beta$.
Thus, as follows from (\ref{e5}), (\ref{e6}), in the low energy limit the
effective action
$W^{CS}[g_{\mu\nu},\beta]$ can be written as
\begin{equation}\label{e7}
W^{CS}[g_{\mu\nu},\beta]={\beta \over \beta_H}W[g_{\mu\nu}]
+ U(\beta)-{\beta \over \beta_H} U(\beta_H)+
\sum_{s}2\pi\xi_s(1-\beta/\beta_H)\int_{\Sigma}d\sigma
<\hat{\phi}^2_s>_\beta
\end{equation}
where $W[g_{\mu\nu}]$ is the action (\ref{i9}) at $\beta=\beta_H$ and
\begin{equation}\label{e8}
U(\beta)=-g(m^2_i){\pi \over 6}{\beta_H \over \beta}
{\cal A}^H~~~,
\end{equation}
\begin{equation}\label{e8a}
g(m^2_i)\equiv
{1 \over 16\pi^2}\left[\sum_s m_s^{D-2}
{}~\Gamma\left(1-\frac D2,m_s^2\delta\right)+2
\sum_d m_d^{D-2}
\Gamma\left(1-\frac D2,m_d^2\delta\right)\right]~~~,
\end{equation}
\begin{equation}\label{e8b}
<\hat{\phi}^2_s>_\beta~\simeq
{1 \over 16\pi^2} m_s^{D-2}
{}~\Gamma\left(1-\frac D2,m_s^2\delta\right)~~~.
\end{equation}
Both quantities $U(\beta)$ and $<\hat{\phi}^2_s>_\beta$ are ultraviolet
divergent. However it is possible to demonstrate that in the induced gravity,
when conditions (\ref{i3})-(\ref{i6'}) are satisfied, the only divergent terms
that enter
the off-shell action $W^{CS}[g_{\mu\nu},\beta]$  are of the second and higher
order in the deficit angle.
These divergences vanish when $\beta=\beta_H$, so that
the corresponding terms do not contribute to the entropy.
The entropy in the conical singularity method (\ref{e1}) reads
\begin{equation}\label{e9}
S^{CS}(\beta_H)=\left.\left(\beta {\partial \over \partial
\beta}-1\right)U(\beta)\right|_{\beta=\beta_H}
-\sum_{s}2\pi\xi_s\int_{\Sigma}d\sigma <\hat{\phi}^2_s>~~~.
\end{equation}
Here $<\hat{\phi}^2_s>\equiv<\hat{\phi}^2_s>_{\beta_H}$ is the average in the
Hartle-Hawking state.
Now it is easy to show by making use of (\ref{i14}) and (\ref{e8}) that the
off-shell induced entropy $S^{CS}$ is ultraviolet
finite at $\beta=\beta_H$ and coincides exactly to the
Bekenstein-Hawking entropy $S^{BH}$:
\begin{equation}\label{e10}
S^{CS}(\beta_H)={1 \over 4G}{\cal A}^H=S^{BH}
\end{equation}
where $G$ is determined by Eq. (\ref{i10}) and ${\cal A}^H$
is the area of the horizon surface.

\section{Relation to statistical mechanics}
\setcounter{equation}0

In the considered model the gravity appears as the collective effect related to
averaging over the fields with the planckian masses. Let us find the total
statistical-mechanical entropy $S^{SM}$ of these ultra heavy constituents and
its relation to $S^{BH}$.  The
statistical-mechanical entropy
\begin{equation}\label{r1}
S^{SM}=-\mbox{Tr}~\hat{\rho}\ln \hat{\rho}~~~
\end{equation}
is determined in terms of the thermal density matrix
\begin{equation}\label{r2}
\hat{\rho}={e^{-\hat{H}/T} \over \mbox{Tr}~e^{-\hat{H}/T}}
\end{equation}
where $\hat{H}$ is the Hamiltonian of the system.
To find $S^{SM}$ for the fields of our model we shall rewrite
(\ref{r1}) in a more suitable form
\begin{equation}\label{r3}
S^{SM}=\left(\beta {\partial \over \partial \beta}-1\right)
W^{SM}
\end{equation}
in terms of the quantity $W^{SM}$, related to the free energy ${\cal F}$ of the
system
\begin{equation}\label{r4}
W^{SM}=T^{-1}{\cal F}~~~,~~~e^{-{\cal F}/T}\equiv\mbox{Tr}~e^{-\hat{H}/T}
{}~~~.
\end{equation}
In (\ref{r3}), as before, $\beta=g_{\tau\tau}^{-1/2}(r_B)T^{-1}$.
For  quantum fields on a static curved background  $W^{SM}$ has the
functional integral representation \cite{DeAlwis} similar to that of the
covariant effective action $W[g_{\mu\nu}]$:
\begin{equation}\label{r5}
\exp(-W^{SM}[g_{\mu\nu},\beta])=\int {\cal D}_H\varphi_i
\exp(-I[\varphi_i,g_{\mu\nu}])~~~.
\end{equation}
We call $W^{SM}$ the statistical-mechanical action.
The difference between (\ref{1.3}) and (\ref{r5}) is in the integration
measures. The statistical-mechanical action is determined with the help of the
non-covariant canonical measure
${\cal D}_H\phi$. The equation (\ref{r5}) can be obtained
from the canonical representation of $\mbox{Tr}~e^{-\hat{H}/T}$ after the
integration over the canonical momenta, see the details in Ref.
\cite{DeAlwis}. The covariant ${\cal D}\phi$ and canonical ${\cal D}_H\phi$
measures are related to each other in the simple way
\begin{equation}\label{r6}
\left. {\cal D}_H\varphi_i\right|_{g_{\mu\nu}}=\left.
{\cal D}\tilde{\varphi}_i\right|_{\tilde{g}_{\mu\nu}}
\end{equation}
where
\begin{equation}\label{r7}
\tilde{g}_{\mu\nu}={g_{\mu\nu} \over g_{\tau\tau}}~~~,~~~
\tilde{\phi}_s=(g_{\tau\tau})^{{D-2 \over 4}}\phi_s~~~,~~~
\tilde{\psi}_d=(g_{\tau\tau})^{{D-1 \over 4}}\psi_d~~~.
\end{equation}
So by making use of the field redefinitions  in (\ref{r5}) one can also rewrite
$W^{SM}$ as a
covariant action \cite{DeAlwis},\cite{Zerbini}
\begin{equation}\label{r8}
W^{SM}[g_{\mu\nu},\beta]=\widetilde{W}[\tilde{g}_{\mu\nu},\beta]
\end{equation}
but for the ultrastatic space $\widetilde{\cal M}$ with the metric
$\tilde{g}_{\mu\nu}$.

A stationary black hole is a special case of the stationary
geometry that possesses the Killing horizon where $g_{\tau\tau}=0$.
In the Euclidean formulation the Euclidean horizon is formed by the fixed
points of the Killing vector field.
In the presence of the horizon the standard canonical
formulation of the theory meets difficulties. In particular, the canonical
integration measure diverges when approaching the horizon.  It means that on a
black hole background the statistical-mechanical quantities are ill defined.
There are different
ways to overcome this difficulty. We use here the volume cut-off prescription
(see, for instance Ref.\cite{FFZ}).
In this method one simply cuts the spatial integrations in the
statistical-mechanical quantities near the horizon at some proper distance
$\epsilon$. As the result $\epsilon$ appears in all physical quantities as an
additional volume cut-off parameter.

Let us calculate now $\widetilde{W}[\tilde{g}_{\mu\nu},\beta]$. This functional
has the form
\begin{equation}\label{r9}
\widetilde{W}[\tilde{g}_{\mu\nu},\beta]=
\sum_{s}\widetilde{W}_s[\tilde{g}_{\mu\nu},\beta]+\sum_{d}
\widetilde{W}_d[\tilde{g}_{\mu\nu},\beta]~~~,
\end{equation}
where the actions of the scalar and Dirac particles read
\begin{equation}\label{r10}
\widetilde{W}_i[\tilde{g}_{\mu\nu},\beta]=-{\eta_i \over 2}
\int_{0}^{\infty}{dt \over t}
\sum_{l=-\infty}^{\infty} e^{-4\pi^2 (l+w_i)^2 \beta^{-2}t}~
\mbox{Tr}~e^{-(L_i+m_i^2g_{\tau\tau}) t}~~,~~i=s,d~~~.
\end{equation}
This formula follows from the structure  of the background space
$\widetilde{\cal M}=S^1\times
\widetilde{\cal M}'$.
The numbers $w_s=0$ and $w_d=1/2$ are related to the different periods in
$\tau$ for integer and half-odd-integer spins. The operators $L_i$ act on
$(D-1)$-dimensional space $\widetilde{\cal M}'$ and their form can be easily
found
by applying the transformation (\ref{r7}) to the operators (\ref{i13}). The
functionals $\widetilde{W}_i[\tilde{g}_{\mu\nu},\beta]$ are ultraviolet
divergent and for their calculation one must introduce the ultraviolet
regulator as well. Let us note that the regularization parameters in the
covariant and
statistical-mechanical actions can be different in some regularization
procedures.  We prefer to  work further with the dimensional regularization in
the parameter $D$. This prescription is identical for the both functionals and
enables us to compute these quantities on the equal footing. This will be
important for us later. The features of this regularization and its
alternatives are discussed below.

It is suitable to use the Poisson formula to rewrite the sums
\begin{equation}\label{r11}
\sum_{l=-\infty}^{\infty} e^{-4\pi^2 (l+w_i)^2 \beta^{-2}t}
={\beta \over \sqrt{4\pi t}}
\sum_{k=-\infty}^{\infty}\eta_i^k e^{-{\beta^2 k^2 \over 4t}}~~~.
\end{equation}
After that the actions can be separated onto the vacuum ($\widetilde{W}_i^V$)
and thermal ($\widetilde{W}_i^T$) parts:
\begin{equation}\label{r12}
\widetilde{W}_i=\widetilde{W}_i^V+\widetilde{W}_i^T~~~,
\end{equation}
\begin{equation}\label{r13}
\widetilde{W}_i^V=-{1 \over 2} \int_{0}^{\infty}{dt \over t}
{\beta \over \sqrt{4\pi t}}~
\mbox{Tr}~e^{-(L_i+m_i^2g_{\tau\tau}) t}~~~,
\end{equation}
\begin{equation}\label{r14}
\widetilde{W}_i^T=-\int_{0}^{\infty}{dt \over t}
{\beta \over \sqrt{4\pi t}}
\sum_{k=1}^{\infty}\eta_i^{k+1} e^{-{\beta^2 k^2 \over 4t}}
\mbox{Tr}~e^{-(L_i+m_i^2g_{\tau\tau}) t}~~~.
\end{equation}
Obviously, $\widetilde{W}_i^V$ are proportional to $\beta$ and give a
contribution only to the vacuum energy, while the entropy is determined by the
thermal parts $\widetilde{W}_i^T$.

Our aim is to calculate the functionals $\widetilde{W}_i^T$ by decomposing them
in powers of the curvature of the physical background ${\cal M}$, similar to
the
decomposition of the covariant action, and then to find from this expansion the
entropy $S^{SM}$. To do this we use the asymptotic series of the heat kernels
of the operators $L_i+m_i^2g_{\tau\tau}$. As we will see it is sufficient to
consider only the zero-order  approximation:
\begin{equation}\label{r15}
\mbox{Tr}~e^{-(L_i+m_i^2g_{\tau\tau}) t}\simeq {n_i \over (4\pi t)^{{D-1 \over
2}}}
\int_{\widetilde{\cal M}'}d\tilde{V}' e^{-m_i^2g_{\tau\tau}t}~~~,
\end{equation}
where $n_s=1$ and $n_d=4$ are the number of components of a field in question
and $d\tilde{V}'$ is the $(D-1)$-volume element on $\widetilde{\cal M}'$. Then
by taking into account that
\begin{equation}\label{r16}
\int^{\beta}_{0}d\tau\int_{\widetilde{\cal M}'}d\tilde{V}'
=\int_{\cal M}dV(g_{\tau\tau})^{-D/2}~~~,
\end{equation}
we get from Eq. (\ref{r14})
the expression
\begin{equation}\label{r17}
\widetilde{W}_i^T\simeq-\int_{0}^{\infty}{dt \over t}
{n_i \over (4\pi t)^{D/2}}
\sum_{k=1}^{\infty}\eta_i^{k+1} e^{-{\beta^2 k^2 \over 4t}}
\int_{\cal M}dV(g_{\tau\tau})^{-D/2} e^{-m_i^2g_{\tau\tau}t}~~~.
\end{equation}
It is written in terms of geometrical characteristics of the physical space
${\cal M}$ with the volume cutoff in the integrals at the proper distance
$\epsilon$ (in the metric $g_{\mu\nu}$)
near the horizon $\Sigma$.

To evaluate the last integral in (\ref{r17}) let us point out that the main
contribution here comes out from the region near $\Sigma$ where $g_{\tau\tau}$
is small. In this region the black hole metric can be approximated by the
Rindler-like metric
\begin{equation}\label{r18}
ds^2=\left({2\pi \over \beta_H}\right)^2y^2d\tau^2+dy^2+dl^2~~,~~0\leq\tau\leq
\beta~~,~~y\geq \epsilon~~~,
\end{equation}
where $dl^2$ is the line element on the horizon.
Then the integration over ${\cal M}$ is reduced to the
surface integral
\begin{equation}\label{r19}
\int_{\cal M}dV(g_{\tau\tau})^{-D/2} e^{-m_i^2g_{\tau\tau}t}=
{\beta\beta_H \over 4\pi}(m_i^2t)^{\frac D2-1}\Gamma\left(1-\frac D2,~
\epsilon^2m_i^2\left({2\pi \over \beta_H}\right)^2t\right){\cal A}^H~~~,
\end{equation}
where $\Gamma(z,\sigma)$
is incomplete gamma function (\ref{i19}). Now we  use the  relation between
complete and incomplete gamma functions:
\begin{equation}\label{r20}
\Gamma(z,\sigma)=\Gamma(z)-\sum_{p=0}^{\infty}{(-1)^p \over p!}
{\sigma^{z+p} \over z+p}
\end{equation}
that is valid when $z\neq 0, -1,-2,..~$. By substituting (\ref{r19}) into
(\ref{r17}) and using (\ref{r20}) one can get
\begin{equation}\label{r21}
\widetilde{W}_i^T=-{\beta\beta_H \over (4\pi)^{D/2+1}}
\Gamma\left(1-\frac D2\right)n_im_i^{D-2}
\left[\int_{0}^{\infty}{dt \over t^2}
\sum_{k=1}^{\infty}\eta_i^{k+1}
e^{-{\beta^2 k^2 \over 4t}}\right]
{\cal A}^H+P_i(\epsilon,D)~~~.
\end{equation}
The function $P_i(\epsilon,D)$ includes all the dependence on the cut-off
parameter $\epsilon$ and has the representation
\begin{equation}\label{r21a}
P_i(\epsilon,D)={\epsilon^{2-D} \over (4\pi)^{D/2}}~
\left({\beta_H \over \pi\beta}\right)^{D-1}Q_i~{\cal A}^H~~~,
\end{equation}
\begin{equation}\label{r21ab}
Q_i=\sum_{p=0}^{\infty}{ (-1)^p \over p!}
\left({\pi\beta \epsilon m_i\over \beta_H}\right)^{2p}~{\zeta_i(D-2p)
\Gamma(D/2-p) \over p+1-D/2}~~~,
\end{equation}
\begin{equation}\label{r21b}
\zeta_i(z)=\sum_{k=1}^{\infty}\eta_i^{k+1}k^{-z}~~~.
\end{equation}
Eqs.(\ref{r21})-(\ref{r21b}) enable one to analyse the behavior of the thermal
part $\widetilde{W}_i^T$ in two different limits.
First one can keep $\epsilon\neq 0$ and remove the ultraviolet regularization
by going to $D=4$. As it should be, the functional $\widetilde{W}_i^T$ doesn't
have the ultraviolet divergences because the pole terms $\sim (D-4)^{-1}$ in
the first term in the r.h.s. in (\ref{r21})
and in $P_i(\epsilon,D)$ cancel each other. However the part $P_i(\epsilon,D)$
develops another divergence in $D=4$ when $\epsilon\rightarrow 0$:
\begin{equation}\label{r22}
P_s\simeq-{1 \over 1440\pi} \left({\beta_H \over \beta}\right)^3{1 \over
\epsilon ^2}~
{\cal A}^H~~~,~~~P_d\simeq n_d~\frac 78~ P_s~~~.
\end{equation}
It is easy to see that $P_s$ and $P_d$ reproduce
the well-known high temperature behavior of the free energy of the relativistic
Bose and Fermi gas.

For our purpose we need another way of computations when one  takes the limit
$\epsilon\rightarrow 0$ first but keeps the ultraviolet regularizator fixed.
The ultraviolet infinities
disappear when $\mbox{Re}D < 0$. If this condition is satisfied the functions
$P_i$ vanish at $\epsilon=0$.
So by integrating the first term in r.h.s. in
(\ref{r21}) we get when $\mbox{Re}D < 0$:
\begin{equation}\label{r23}
\widetilde{W}_s^T=-{\pi \over 6}{\beta_H \over \beta}
{\Gamma(1-D/2) \over (4\pi)^{D/2}}m_s^{D-2}{\cal A}^H~~~,
\end{equation}
\begin{equation}\label{r24}
\widetilde{W}_d^T=-{\pi \over 3}{\beta_H \over \beta}
{\Gamma(1-D/2) \over (4\pi)^{D/2}}m_d^{D-2}{\cal A}^H~~~.
\end{equation}
Finally substituting (\ref{r23}) and (\ref{r24}) we get the following
expression for the statistical-mechanical action:
\begin{equation}\label{r25}
W^{SM}=\sum_i\widetilde{W}_i\simeq\sum_i\widetilde{W}_i^V+
U(\beta)~~,~~\mbox{Re}D < 0~~~,~~~\epsilon=0~~,
\end{equation}
where $U(\beta)$ is defined by expressions (\ref{e8}) and (\ref{e8a})
calculated in the dimensional regularization scheme ($\delta=0$).
This equation represents rather interesting result because it demonstrates that
statistical-mechanical action includes
the same function $U(\beta)$ which appears in the covariant action on the
singular instanton, see (\ref{e7}).  This is the function which, being
regularized according to our prescription, determines the
statistical-mechanical entropy
(\ref{r3}) of the heavy constituents:
\begin{eqnarray}\label{r26}
S^{SM}=\left.\left(\beta {\partial \over \partial
\beta}-1\right)U(\beta)\right|_{\beta=\beta_H}~~~.
\end{eqnarray}
Let us suppose now that the field average (\ref{e8b}) is calculated in the same
regularization. Then, according to (\ref{e10}), we can rewrite the
Bekenstein-Hawking entropy in the following form
\begin{eqnarray}\label{r27}
S^{BH}&=&\left.S^{CS}\right|_{\beta=\beta_H}=
S^{SM}-\sum_{s}2\pi\xi_s\int_{\Sigma}d\sigma<\hat{\phi}^2_s>  \\
&=& - \mbox{\bf Tr}~[ \hat{\rho} \ln \hat{\rho} ]  -
\sum_{s}2\pi\xi_s\int_{\Sigma}d\sigma<\hat{\phi}^2_s> \nonumber
\end{eqnarray}
It is assumed that the regularization here should be removed
at the end, simultaneously in $S^{CS}$ and in $<\hat{\phi}^2_s>$.
This relation is our main result.

We see therefore that the part of the black hole entropy in the induced gravity
is the statistical-mechanical entropy of the
heavy constituents with the Planckian masses. The other part,
that is expressed in terms of  fluctuations of scalar fields on the horizon, is
not related (at least in our model) to the counting of states. The role of the
horizon term is to remove the divergences
in $S^{SM}$. It is easy to see that $S^{BH}$ cannot be represented as a pure
statistical-mechanical entropy of the
thermal gas around a black hole, because all fields, regardless the spin, give
the positive divergent contributions into $S^{SM}$. So the
statistical-mechanical representation of
$S^{BH}$ requires a subtraction procedure. There is some similarity between
this subtraction procedure and one discussed in Refs.\cite{FFZ},\cite{FFZ2}.

It should be also emphasized that because of the exponential decrease of the
Euclidean heat kernels the main contributions
to the integrals (\ref{r19}) appearing in $\widetilde{W}_i^T$ are given by the
region close to the horizon where $g_{\tau\tau}\ll 1$. In other words,
in the induced gravity the entropy $S^{SM}$ comes from the narrow layer located
near the horizon and having the depth of order of the Planck length $\sim
m_{Pl}^{-1}$. Hence  $S^{SM}$ depends only on the local properties of the
horizon. This conclusion implies that the result
(\ref{r27}) must be valid for a general static or stationary
black hole in the Einstein theory and in more general induced gravity theories.

A remark about the regularization scheme is in order. To write
the black hole entropy in the form (\ref{r27}) the quantities
$S^{SM}$ and $<\hat{\phi}^2_s>$ must be calculated by the same method. The
dimensional regularization used here is the easiest tool to do this. However,
as it is known, such regularization does
not take into account some of the ultraviolet divergences.
An alternative way could be to use the Pauli-Villars regularization based on
the introduction of the fictitious particles with the wrong statistics
\cite{Myers}. The Pauli-Villars scheme is more complicated but one can verify
that it gives the same result for $S^{BH}$.
This indicates that Eq. (\ref{r27}) does not depend on the regularization.
Another observation is that for $\xi_s=0$ the statistical-mechanical  entropy
(\ref{r26}) obtained for the scalar fields in the Pauili-Villars regularization
coincides
with the result for the entropy
derived in \cite{Myers} in the WKB approximation directly from the spectrum of
the energy operator.

 \section{Conclusions}

In conclusion, we would like to stress that all our calculations were
made for a special class of models of induced gravity.  The main
point of our consideration is construction and comparison of the
covariant $W$ and statistical-mechanical $W^{CS}$ effective actions.
The functional  $W$ generates the Einstein action as the low-energy
limit of the induced theory, while the functional $W^{CS}$, being
presented in (3+1)-form, allows one to relate the entropy to the
statistical properties of the constituents. Moreover, with the help
of $W^{CS}$  thermodynamical characteristics of a black hole can be
presented in the form of spatial integrals of local quantities, in
which only narrow region $\sim m^{-1}_{Pl}$ near the horizon does
contribute.

The mechanism relating $S^{BH}$ to the statistical-mechanics of
constituents seems to be quite general and it should work for
the theories of different types. Any two microscopically different 
theories that
result in the same low-energy limit for induced gravity  must predict
the same value of the black hole entropy. 
The details of the statistical-mechanical calculations and the form
of the representation of the black hole entropy in terms of
constituents may differ, but the final result (at least for black
holes of mass much greater that $m_{Pl}$) must be 
determined only by the
form of the low energy effective action for gravity and 
macroscopic parameters of a black hole. This requirement
of consistency of the statistical mechanics of constituents
in fundamental theory with the standard low energy gravitational
calculations can be formulated as the general principle, that we call
the {\em low-energy censorship conjecture}.

In particular this conjecture implies that  in  superstring
theory, which induces gravity as the effective low-energy theory, and
where the metric $g_{\mu\nu}$ arises as the result of collective
string excitations the statistical-mechanical calculations for string
constituents must  reproduce the Bekenstein-Hawking entropy. The
recent calculations of the black hole entropy obtained in the
superstring theory for the special type black hole solutions (see,
for instance, \cite{strings1}-\cite{strings3} and references therein)
might be considered as supporting this point of view.

\vspace{12pt}
{\bf Acknowledgements}:\ \ This work was supported  by the Natural
Sciences and
Engineering
Research Council of Canada.


\begin{thebibliography}{000}
\bibitem{Beke:72} J.~D.Bekenstein, Nuov. Cim. Lett. {\bf 4}   (1972)
737; Phys. Rev. {\bf D7} (1973) 2333;  Phys. Rev. {\bf D9}  (1974)
3292.
\bibitem{Hawk:75} S.W.Hawking, Comm. Math. Phys. {\bf 43} (1975) 199.
\bibitem{GiHa:76} G.~W.~Gibbons and S.~W.~Hawking,  Phys. Rev. {\bf
D15} (1976) 2752.
\bibitem{Hoof:85} G.'t Hooft, Nucl. Phys. {\bf B256} (1985) 727.
\bibitem{Sorkin} L. Bombelli, R. Koul, J. Lee, and R. Sorkin,
Phys. Rev. {\bf D34} (1986) 373.
\bibitem{Srednicki} M. Srednicki, Phys. Rev. Lett. {\bf
71} (1993) 666.
\bibitem{FroNo} V. Frolov and I. Novikov, Phys. Rev. {\bf D48} (1993) 4545.
\bibitem{Frol:95} V.P. Frolov, Phys. Rev. Lett. {\bf 74} (1995) 3319.
\bibitem{SU} L. Susskind and J. Uglum, Phys. Rev. {\bf D50} (1994) 2700.
\bibitem{CW} C. Callan and F. Wilczek, Phys. Lett. {\bf B333}
(1995) 55.
\bibitem{FS} D.V. Fursaev and S.N. Solodukhin, Phys. Lett. {\bf B365} (1996)
51.
\bibitem{Myers} J.-G. Demers, R. Lafrance, and R.C. Myers,
Phys. Rev. {\bf D52}  (1995) 2245.
\bibitem{DeAlwis} S.P. deAlwis and N. Ohta, Phys. Rev. {\bf D52}
(1995) 3529.
\bibitem{Solod} S.N. Solodukhin, Phys. Rev. {\bf D52} (1995) 7046.
\bibitem{Zerbini} A.A. Bytsenko, G. Cognola and S. Zerbini,
Nucl. Phys. {\bf B458} (1996) 267.
\bibitem{Kabat} D. Kabat, Nucl. Phys.
{\bf B453} (1995) 281.
\bibitem{LW} F. Larsen and F. Wilczek, Nucl. Phys. {\bf B458} (1996) 249.
\bibitem{FFZ} V.P. Frolov, D.V. Fursaev and A.I. Zelnikov,
{\it Black Hole Entropy: Off-shell vs On-shell},
preprint hep-th/9512184, to appear in Phys. Rev. {\bf D}.
\bibitem{Mann} R. B. Mann and S. N. Solodukhin, {\it Conical Geometry
and Quantum Entropy of Charged Black Hole}, preprint WATPHYS-TH-96-04,
hep-th/9604118
\bibitem{Zerbini2} G. Cognola, S. Zerbini and L. Vanzo,
{\it Euclidean Approach to the Entropy for a Scalar Field
in Rindler-like Space-Times}, preprint UTF-372, hep-th/9603106.
\bibitem{BaSol} A.D. Barvinsky and S.N. Solodukhin, {\it
Nonminimal Coupling, Boundary Terms and Renormalization
of the Einstein-Hilbert Action and Black Hole Entropy},
preprint WATPHYS-TH-95-11,
gr-qc/9512047.
\bibitem{FFZ2} V.P. Frolov, D.V. Fursaev and A.I. Zelnikov,
{\it Black Hole Entropy: Thermodynamics, Statistical Mechanics
and Subtraction Procedure}, preprint hep-th/9603175,
to appear in Phys. Lett. {\bf B}.
\bibitem{Jacobson} T. Jacobson, {\it Black Hole Entropy and Induced Gravity},
preprint gr-qc/9404039.
\bibitem{Sakh} A.D. Sakharov, Sov. Phys. Doklady, {\bf 12}
(1968) 1040.
\bibitem{York:86} J.~W.~York,  Phys. Rev. {\bf D33} (1986) 2092.
\bibitem{BBWJ} H.~W.~Braden, J.~D.~Brown, B.~F.~Whiting, and J.~W.~Jork, Phys.
Rev. {\bf D42} (1990) 3376.
\bibitem{DF} D.V. Fursaev, Class. Quantum. Grav. {\bf 11}
(1994) 1431.
\bibitem{CKV} G. Cognola, K. Kirsten and L. Vanzo, Phys. Rev.
{\bf D49} (1994) 1029.
\bibitem{FM96} D.V. Fursaev and G. Miele, {\it Cones, Spins
and Heat Kernels}, preprint ALBERTA-THY-17-96,
hep-th/9605153.
\bibitem{strings1} F. Larsen and F. Wilczek, Phys. Lett.
{\bf B375} (1996) 37.
\bibitem{strings2} F. Larsen and F. Wilczek, {\it Classical Hair in String
Theory}, preprint PUPT-1614, hep-th/9604134.
\bibitem{strings3} G.T. Horowitz, J.M. Maldacena,
A. Strominger, {\it Nonextremal Black Hole Microstates and
$U$ Duality}, preprint hep-th/9603109.

\end{thebibliography}
\end{document}